\documentclass[twocolumn]{revtex4}
\usepackage{amssymb,epsfig,amscd}
\usepackage{graphicx}
\usepackage{dcolumn}
\usepackage{bm}
  
\usepackage{epsfig}

\newcommand{\dirac}{\partial\llap{$\diagup$\kern-2pt}}

\def\be{\begin{equation}} 
\def\ee{\end{equation}}
\def\bq{\begin{eqnarray}} 
\def\eq{\end{eqnarray}}

\begin{document}

\title{Phase transition from nuclear matter to color superconducting quark matter: the effect of the isospin}
\author{G.~Pagliara, J.~Schaffner-Bielich}

\affiliation{Institut f\"{u}r Theoretische Physik, Ruprecht-Karls-Universit\"at,
   Philosophenweg 16,  D-69120, Heidelberg, Germany}

\begin{abstract} 
We compute the mixed phase of nuclear matter and 2SC matter for
different temperatures and proton fractions. After showing that the
symmetry energy of the 2SC phase is, to a good approximation, three
times larger than the one of the normal quark phase, we discuss and
compare all the properties of the mixed phase with a 2SC component or
a normal quark matter component.  In particular, the local isospin
densities of the nuclear and the quark component and the stiffness of
the mixed phase are significantly different whether the 2SC phase or
the normal quark phase are considered. If a strong diquark pairing is
adopted for the 2SC phase, there is a possibility to eventually enter
in the nuclear matter 2SC matter mixed phase in low energy heavy ions
collisions experiments.  Possible observables able to discern between
the formation of the 2SC phase or the normal quark phase are finally
discussed.
\end{abstract}

\maketitle

\section{Introduction}
The phase transition from nuclear (hadronic) matter to quark matter at
high density may depend strongly on the isospin asymmetry of
matter. While the value of the nuclear symmetry energy is known at
saturation, its behavior at densities larger than nuclear density is
still under study both theoretically and experimentally by means of heavy ions collisions experiments at low energy
\cite{Steiner:2004fi,Li:2008gp}.  On the other hand, very little is
known about the dependence on the isospin asymmetry of the energy of
the quark phase which is believed to take place at large
density. There have been studies on the phase transition from nuclear
matter to quark matter at large isospin densities mainly focused on
the role of the nuclear matter symmetry energy while the interactions
between quarks have been neglected or limited to the perturbative QCD
corrections \cite{Muller:1997tm,DiToro:2006pq,
Bonanno:2007kh,Bonanno:2008tt,DiToro:2009ig}. 
The main result of these studies is a steep
reduction of the critical density for the phase transition as the
isospin asymmetry increases. This is due to the large value of the
symmetry energy of nuclear matter and the small value of the symmetry
energy of quark matter which is provided only by the Fermi kinetic
contribution (eventually corrected by the perturbative QCD
interactions). Moreover, it was argued in \cite{DiToro:2006pq} that
the large difference between the symmetry energy of the nuclear and
the quark phase could induce, within the mixed phase, the so called
neutron distillation effect: the quark component of the mixed phase is
much more isospin asymmetric with respect to the nuclear phase. In
turn, this effect could modify the particles yield ratios and could be
eventually detected in heavy ions experiments at low energies as the
ones planned at FAIR and NICA.

An important effect was not considered in the above mentioned studies:
at large densities quark matter is likely to be in a color
superconducting state \cite{Alford:2007xm} with superconducting gaps
ranging from a few tens of MeV up to $200$ MeV if a strong diquark
pairing is adopted \cite{Ruester:2005jc,Blaschke:2005uj}. Thus a color
superconducting state could potentially survive even at temperatures
of the order of $50$ MeV or more and it is therefore interesting to
investigate its possible formation in low energy heavy ions
experiments
\footnote{In the so called Polyakov loop NJL model (PNJL), the
critical temperature is higher than the one obtained in the NJL model,
thus within this model the formation of the 2SC phase at high
temperature is even more likely
\cite{Roessner:2006xn,GomezDumm:2008sk}.}. Also in the case in which
the temperature reached in the collision is higher than the critical
temperature for color superconductivity interesting precursor phenomena of the
diquark formation might take place \cite{Kitazawa:2001ft,Voskresensky:2004jp}.

Among the many possible color superconducting phases, the 2SC phase is
the relevant candidate for heavy ions collisions, since it is likely
to appear at lower densities with respect to the three flavor CFL
phase and because, at low collision energy, a small amount of strange
quarks is produced. Based on general arguments, one could expect a
different value of the symmetry energy of the 2SC phase with respect
to the normal quark phase since the Cooper pairs are in a isospin
symmetric state, i.e. the density of up and down quarks are forced to
be equal to allow the formation of the diquark condensate
\cite{Rajagopal:2000ff}.  On the other hand at large isospin asymmetries
the 2SC pairing pattern is broken and the normal quark phase is
obtained as shown in previous papers
\cite{Bedaque:1999nu,Kiriyama:2001ud,Huang:2002zd,Huang:2003xd,Shovkovy:2003uu,Reddy:2004my}.

The aim of this paper is to extend previous calculations on the effect
of isospin on the nuclear matter - quark matter phase transition at
high density and finite temperature by including the non-perturbative
interactions between quarks responsible for the phenomenon of color
superconductivity in the 2SC phase. We also discuss under which
conditions this state of matter might be created in the laboratory and
its possible observational signatures in the framework of the search
for the nuclear matter - quark matter mixed phase in heavy ions experiments
\cite{Sissakian:2006dn}.

The paper is organized as follows: in section II we describe the model
adopted to calculate the 2SC equation of state and we compute its
symmetry energy and compare it with the one of normal quark matter.
In section III we compute the critical densities for the onset of the
phase transition at different asymmetries and temperatures and we
explain the features of the mixed phases.  In section IV we present
the phase diagrams. Further discussions and conclusions are given in
section V.

\section{Quark matter and nuclear matter equations of state}

The region of the QCD phase diagram we want to investigate, baryon
density $n_B$ from nuclear matter density, $n_0=0.16$fm$^{-3}$, to $3-4
n_0$ and temperature T from 0 to $\sim 100$ MeV has a very rich
structure: at these conditions the chiral phase transition is believed
to occur and also a ``deconfinement'' phase transition, here just
meant to be a change from hadronic degrees of freedom to quarks
degrees of freedom. These two transitions are not necessarily coincident,
for instance it has been proven, in the large $N_c$ limit, the existence of a phase in which
chiral symmetry is restored but quarks are 
confined, the so called Quarkyonic phase \cite{McLerran:2007qj,Andronic:2009gj}.
Moreover the phase transition could involve Normal
Quark matter (NQ) or color superconducting matter, the 2SC phase,
depending on the temperature, the isospin asymmetry and the value of
the superconducting gap.  Needless to say, at these regimes QCD is
strongly non-perturbative and one has to resort to models to obtain
some qualitative results.  One possible approach is to consider quark
chiral models, like the NJL or the PNJL models, and to compute the
structure of the phase diagram by means of order parameters as
the chiral condensates, the diquark condensates
\cite{Toublan:2003tt,Frank:2003ve,Barducci:2004tt,Buballa:2003qv,Ruester:2005jc,Blaschke:2005uj} and the Polyakov loop
\cite{Roessner:2006xn,GomezDumm:2008sk}. One important effect missing
in these calculations is confinement. Thus one can predict the line of
the chiral phase transition (or deconfinement phase transition in the
PNJL model) in the temperature chemical potential plane but one has no
estimates for the value of the baryon density or energy density of the
onset of the phase transition which are crucial to make comparisons with the
experiments.

Another approach is to consider two models, one for the low density
low temperature hadronic phase and one for the high density high
temperature quark phase and then to compute the binodal boundaries by
using a Maxwell or a Gibbs construction. Although this approach is
certainly not satisfying because of the use of two different
Lagrangians to describe the same matter, it has the advantages of
providing some numerical estimates of the critical densities and
therefore we will adopt it here.  Some promising studies were already
performed trying to describe the nucleons within a NJL-type model with
quark degrees of freedom but unfortunately the properties of nuclear
matter at saturation cannot be correctly reproduced
\cite{Lawley:2005ru,Rezaeian:2005nm,Lawley:2006ps}. Recently, a new
approach has been proposed in which a unique Lagrangian is considered
having both hadronic and quark degrees of freedom; the phase
transition between nuclear matter and quark matter is regulated by the
Polyakov loop \cite{Dexheimer:2009hi}. While interesting, this model
neglects the formation of diquark condensates.

Let us start by describing the quark model we adopt in our work. In the same
spirit of Refs.~\cite{Alford:2002kj,Alford:2004pf}, we start with the
thermodynamic potential of normal quark matter within the MIT bag
model for two massless flavors, up and down quarks, and we correct it
with the contribution from the quark pairing as calculated in
Refs.~\cite{Huang:2002zd,Huang:2003xd}: 
\be
\Omega=\Omega_{NQ}+\Delta_{2SC}
\ee

where:
\be
\Omega_{NQ}=-\frac{\mu_u^4}{4\pi^2}-\frac{\mu_d^4}{4\pi^2}-\frac{1}{2}T^2 \mu_u^2-\frac{1}{2} T^2 \mu_d^2-\frac{7}{30} \pi^2 T^4 + B
\ee

$\mu_u$ and $\mu_d$ are the chemical potentials of up and down quarks and $B$ is the bag constant. 

We treat now separately $\Delta_{2SC}$ by using the formalism of
Ref.~\cite{Huang:2002zd,Huang:2003xd} in which a NJL like model is
proposed to treat the diquark pairing.  Let us start by introducing
the quark chemical potentials $\mu_{i,\alpha}$ where $i={up,down}$ is
the flavor index and $\alpha={r,g,b}$ is the color index for red,
green and blue quarks.  The relations of chemical equilibrium read:
\bq
\mu_{ur}&=&\mu_{ug}=\mu+\frac{2}{3}\mu_c+\frac{1}{3}\mu_8\\
\mu_{dr}&=&\mu_{dg}=\mu-\frac{1}{3}\mu_c+\frac{1}{3}\mu_8\\
\mu_{ub}&=&\mu+\frac{2}{3}\mu_c-\frac{2}{3}\mu_8\\
\mu_{db}&=&\mu-\frac{1}{3}\mu_c-\frac{2}{3}\mu_8
\eq

where $\mu$, $\mu_c$ and $\mu_8$ are the quark chemical potential, the
charge chemical potential and the chemical potential associated with
the color generator $T_8$ of $SU(3)_c$. The difference between the thermodynamic
potential of quarks in the 2SC phase and of quarks in the normal
phase, in mean field approximation \cite{Huang:2002zd,Huang:2003xd},
reads:

\bq
\Delta_{2SC}&=&-2 \sum_a\int \frac{\mathrm{d}^3 p}{(2 \pi)^3}\left(E_a^{2SC}+2T\mathrm{ln}(1+\exp(\frac{-E_a^{2SC}}{T}))\right. \nonumber \\
&-&\left.E_a^{NQ} -2T\mathrm{ln}(1+\exp(\frac{-E_a^{NQ}}{T})) \right)+\frac{\Delta^2}{4 G_D}
\eq 

$G_D$ is the diquark coupling, $\Delta$ is the superconducting gap,
$E_a^{2SC}$ are the (6 quark and 6 antiquark) quasiparticles dispersion relations
in the 2SC phase as calculated in \cite{Huang:2002zd,Huang:2003xd}:
\bq
E^{\pm}_{ub}&=&p \pm \mu_{ub} \,\,\,\,\,[\times 1]\\
E^{\pm}_{db}&=&p \pm \mu_{db} \,\,\,\,\,[\times 1]\\
E^{\pm}_{\Delta^{\pm}}&=&\sqrt{(p \pm \overline{\mu})^2+\Delta^2} \pm \delta \mu \,\,\,\,\,[\times 2]
\eq

the numbers in the square brackets represent the degeneracy, 
$\overline{\mu}=(\mu_{ur}+\mu_{dg})/2$, $\delta \mu = (\mu_{dg}-\mu_{ur})/2$
and $E_a^{NQ}$ the free quarks (6) and antiquarks (6) dispersion relations
in the normal quark phase (obtained by setting $\Delta=0$,
$\mu_8=0$ in $E_a^{2SC}$). Notice that here we adopt the approximation of massless
quarks and we do not take into account the effective mass of quarks
because, as we will see in the next section, the phase transition from
nuclear matter to quark matter occurs at values of the baryon chemical
potential larger then the chemical potential of the restoration of
chiral symmetry as obtained in the model of
\cite{Huang:2002zd,Huang:2003xd}. The integrals, as usual, are
regularized by introducing a cut-off $\Lambda$.  The values of
$\Delta$ and $ \mu_8$ are obtained by minimizing $\Delta_{2SC}$ with
respect to $\Delta$ and $\mu_8$.  The equation of state can then be
computed as a function of $\mu$, $\mu_c$ and $T$
\footnote{Notice that by subtracting in the expression of
$\Delta_{2SC}$ the free quarks dispersion relations allows to impose
the condition that the thermodynamic potential is $B$ at zero density
and temperature as in the MIT bag model and moreover allows to obtain
the normal quark equation of state in the case in which $\Delta$ and $
\mu_8$ are set to zero.  This would not be the case without this
subtraction scheme because the integrals in the NJL model are
regularized by the cut-off and therefore at large temperatures the
tails of the Fermi distributions are cut away.}.

\begin{figure}
    \begin{centering}
\epsfig{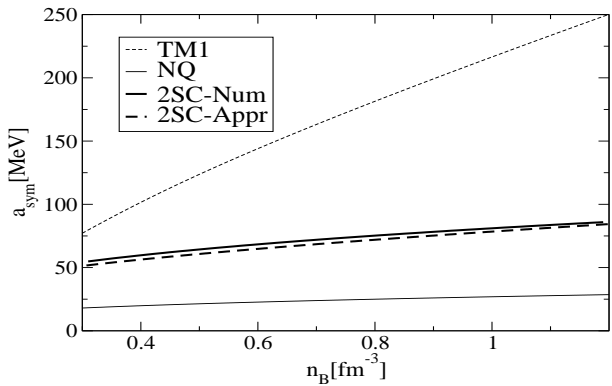}
    \caption{Symmetry energy of the nuclear phase TM1 (thin dashed lines),
2SC phase (numerical calculation, thick solid line), 2SC phase (approximation, thick dashed line) and normal quark phase (thin solid line)
as functions of the density. The symmetry energy of the 2SC phase
is roughly three times larger than the one of the normal quark phase. 
The nuclear phase has the largest value of the symmetry energy. Here, 
the intermediate value of the diquark coupling has been used.
\label{fig:asy} }
   \end{centering}
\end{figure} 

Let us discuss now our choice of the parameters: as in
\cite{Huang:2002zd,Huang:2003xd} $\Lambda=0.6533$ GeV and we consider
two values of the diquark coupling $G_D=3/4G_S$ and $G_D=G_S$, where
$G_S=5.0163$ GeV$^{-2}$ is the quark-antiquark pairing, with
corresponding gaps $\Delta \sim 130$ MeV and $\Delta \sim 200$ MeV for
symmetric matter, at zero temperature and at $\mu=500$ MeV. We will
discuss the values of $B$ in the next section.

Concerning the Nuclear Matter (NM) equation of state, we adopt the widely used
relativistic mean field model with the parameterization TM1
\cite{Shen:1998gq}.  In our calculation we will not include the pion
contribution to the nuclear equation of state. The are two reasons for
which we neglect pions: their contribution to the pressure is
important for large temperatures $\sim 100$ MeV and small chemical
potentials. The second point is that within the relativistic mean
field model we are using, the effective mass of the pion as a function
of the temperature and the density cannot be computed and, by using
its vacuum mass, it is well known that such kind of models for nuclear
matter predicts pion condensation at densities not far from
saturation. The possibility of pion condensation seems, on the other
hand, to be ruled out in more sophisticated chiral models in which
the effective mass of the pion can be computed and it turns out to be
large enough to prevent the condensation \cite{Bonanno:2008tt}.

We want to give now some arguments concerning the behavior of 
the quark equation of state as a function of the isospin asymmetry.
First, we define the asymmetry $t$ for the quark phase as:
\be
t=3\frac{n_d-n_u}{n_d+n_u}=\frac{n_d-n_u}{n_B}
\ee
where $n_d$ and $n_u$ are the densities of down and up quarks and
$n_B$ is the baryon density. The proton fraction is related to $t$ by
the relation $Z/A=(1-t)/2$. In the normal quark phase, since
interactions are neglected, only the Fermi kinetic terms contribute to
the symmetry energy which can be easily shown to be
$a_{sym}^{NQ}=\mu/6=\mu_B/18$.  In the 2SC phase, the formation of
Cooper pairs forces the densities of the paired quarks to be the same.
Only the blue up and down quarks are unpaired and therefore can
eventually have different densities. Thus, to a good approximation, at
fixed values of $\mu$ and $\mu_c$ the asymmetry of the 2SC phase is 
$1/3$ of the asymmetry of the normal quark phase $t^{2SC} \simeq t^{NQ}/3$. The
approximation consists in neglecting the correction to the paired
quark densities given by the gap which in fact scales as
${(\Delta/\mu)}^2$ \cite{Rajagopal:2000ff}. At fixed $\mu$ and $\mu_c$ the total quark densities
of the 2SC phase and the normal quark phase are quite similar. On
the other hand, at fixed density and isospin asymmetry, we expect the
mismatch between the chemical potentials of up and down quarks to be
larger in the 2SC phase than in the normal quark phase (because only
the two blue unpaired quarks contribute to the asymmetry).  We can
write the energy per baryon of the 2SC phase as:
\be
(E/N)^{2SC} \simeq \frac{(E/N)^{NQ}}{3} + \text{paired quarks contribution} 
\ee
the first contribution corresponds to the blue quarks and it is $\simeq 1/3$ of the energy per baryon of the normal quark phase 
(again, by neglecting the correction to the paired quark densities given by the gap).
The symmetry energy turns out to be:
\be
a_{sym}^{2SC}=\frac{1}{2} \frac{\mathrm{d}^2 (E/N)^{2SC}}{(\mathrm{d}t^{2SC})^2} \simeq \frac{1}{2} \frac{\mathrm{d}^2 (E/N)^{NQ}/3}{((\mathrm{d}t^{NQ}/3)^2}=3a_{sym}^{NQ}
\ee

To check this result, we calculate numerically the symmetry energy of the 2SC phase
by using the general definition of symmetry energy:
\be
(E/N)_t = (E/N)_{t=0}+a_{sym}t^2
\ee
from which its value is easily obtained 
by computing the equations of state of the 2SC phase for $t=0$ and for another value
of $t$, $t=0.2$ in our calculation. 

\begin{figure}
    \begin{centering}
\epsfig{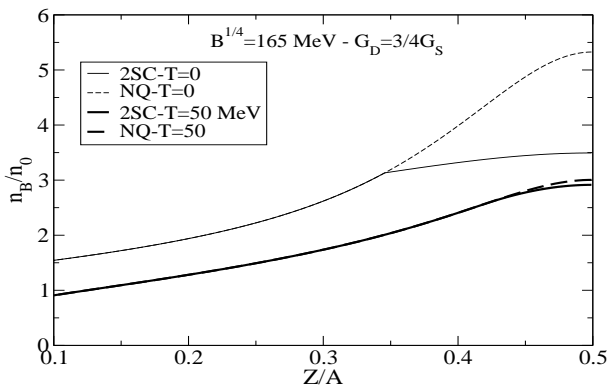}
    \caption{Density for the onset of the phase transition as a function of the proton fraction Z/A and for two values of temperature
$T=0$ and $T=50$ MeV, the intermediate value of the diquark pairing is adopted. The onset of the phase transition
is smaller in the case of the 2SC phase (solid thin and thick lines) with respect to the case of normal quark matter (dashed thin and thick lines).
Before the unpairing transition at $Z/A \sim 0.35$ for $T=0$ and $Z/A \sim 0.42$ for $T=50$ MeV, 
the critical density for the onset of the 2SC phase has a mild dependence on the proton fraction while a strong
dependence is evident in the case of normal quark matter.
\label{fig:1} }
   \end{centering}
\end{figure} 

In Fig.~1 we show the numerical and the approximated results for the symmetry energy of the
2SC phase and the symmetry energy of the normal quark phase as functions of the baryon density. For
comparison we also show the symmetry energy of nuclear matter as
obtained in the relativistic mean field model TM1 (given by the sum of the Kinetic term and the isovector term \cite{DiToro:2006pq}).  
Our approximation for the calculation of the symmetry energy of the 2SC
phase works pretty well, within an error of a few percent it is three times larger than the one of the
normal quark phase. The symmetry energy of nuclear matter is of course
larger than the ones of the quark phases but at densities of the order of
two times saturation density it is actually comparable with the one of
the 2SC phase.  As we will show in the next sections, the larger value
of the symmetry energy of the 2SC phase with respect to the normal
quark phase is responsible for the behavior of the critical density of
the onset of the phase transition as a function of the asymmetry and
it affects also the properties of the mixed phase.

\section{Mixed phases at different Z/A}

The mixed phase between quark matter and nuclear matter is computed by
solving the Gibbs conditions for a multicomponent system with two
globally conserved charges \cite{Glendenning:1992vb,Hempel:2009vp}, the baryonic
charge and the isospin charge associated with the chemical potentials
$\mu_B=\mu_n = 3\mu$ and $\mu_I=(\mu_p-\mu_n)/2=(\mu_d-\mu_u)/2$ respectively, where $\mu_n$ and $\mu_p$
are the chemical potentials of neutrons and protons.  The system can
analogously be described by imposing the global conservation of the
baryonic charge and the electric charge, the electric charge chemical
potential being $\mu_c=\mu_I/2$.  We prefer to work by using the
second description therefore we will use the electric charge ratio or
proton fraction Z/A, and the electric charge chemical potential as the
second conserved quantity in addition to the baryonic charge. The
Gibbs conditions read:

\bq
P^{NM}(\mu_B,\mu_c,T)&=&P^{2SC}(\mu_B,\mu_c,T)\\
Z/A\,n_B &=& (1-\chi)n_c^{NM}+\chi n_c^{2SC}
\eq 
where $P^{NM}$, $P^{2SC}$ are the pressures of the nuclear and the quark phase,
$n_c^{NM}$, $n_c^{2SC}$ are the charge densities of the nuclear and the quark phase, $\chi$ is the volume fraction of the quark phase
and $n_B=(1-\chi)n_B^{NM}+\chi n_B^{2SC}$ is the baryon density.

As a first step, we compute the value of the critical density for the
onset of the nuclear matter - 2SC mixed phase at fixed temperature and
by varying the proton fraction.  Concerning the choice of the bag
parameter, we select two values $B^{1/4}=165$ MeV and $B^{1/4}=190$
MeV which together with the choice of intermediate and strong diquark
pairing, $G_D=3/4 G_S$ and $G_D=G_S$, allow to obtain the onset of the mixed phase a $T=0$ and
for symmetric matter at densities larger than $\sim 3 n_0$, in agreement,
as argued in \cite{Aichelin:2008jn}, with the constraint put by the SIS data.  Results are shown in Figs 2 and 3, where also the case of
normal quark matter is shown for comparison. At zero temperature and
for symmetric matter, the onset of the mixed phase is strongly reduced
in the 2SC phase with respect to the normal phase: this is clearly due
to the softer 2SC equation of state in which the formation of Cooper
pairs allows the system to lower its energy with respect to the case
of a system of unpaired quarks.  As the proton fraction decreases, or
the asymmetry increases, a steep reduction of the critical density
is obtained for normal quark matter as noticed in
Refs.\cite{Muller:1997tm,DiToro:2006pq,DiToro:2009ig} due to the
strong stiffening of the nuclear equation of state. This effect is due to the
large value of the nuclear symmetry energy.  On the other hand, as the
proton fraction is reduced, the normal quark equation of state has a
mild dependence on the asymmetry due to the small value of its
symmetry energy.  In the case of 2SC phase instead, the critical
density stays almost constant as the proton fraction is reduced,
because of the larger value of the symmetry energy of the 2SC phase:
both the nuclear and the 2SC equations of state become substantially
stiffer when the proton fraction is reduced and therefore the critical
density is almost independent of the proton fraction.  As the proton
fraction is further decreased, at some point the stress caused on the
up and down quark's Fermi surfaces becomes too large, the 2SC pairing
is broken and only the normal quark phase can be formed in the phase
transition. Notice the effect of the different values of the diquark
pairing we are adopting: while in Fig.~2, for intermediate pairing, the
unpairing transition is obtained for $Z/A \sim
0.35$ at zero temperature, for the strong pairing case, Fig.~3, this occurs at $Z/A \sim
0.25$.  For extremely low values of Z/A, as the ones reached in neutron
star matter, by computing the mixed phase of
nuclear matter and color superconducting
matter, an interesting effect was obtained in
Ref.~\cite{Blaschke:2008br}: the conditions of beta stability and
charge neutrality render the effective mass of up quarks larger than
the one of down quarks and one obtains first a mixed phase between
nuclear matter and down quarks and only at larger densities also the
up quarks are deconfined. We do not consider here this possibility.

\begin{figure}
    \begin{centering}
\epsfig{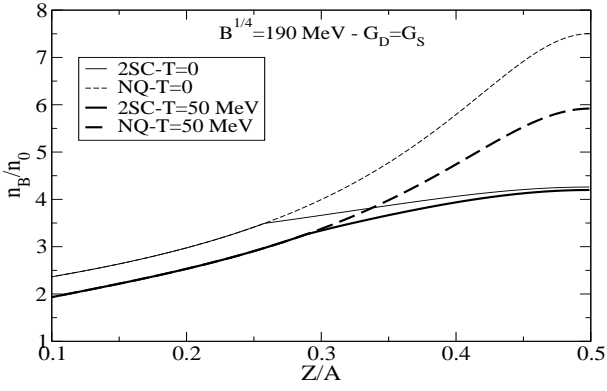}
    \caption{Density for the onset of the phase transition as a function of the proton fraction Z/A and for two values of temperature
$T=0$ and $T=50$ MeV, the strong diquark pairing is adopted (the lines denote the same as in Fig.1). Due to the the large value of the gap,
the unpairing transition occurs at $Z/A \sim 0.25$ for $T=0$ and $Z/A \sim 0.3$ for $T=50$ MeV.
\label{fig:2} }
   \end{centering}
\end{figure}

Finally, the dependence on the temperature is also interesting: at finite
temperature, $T=50$ MeV in Fig.~\ref{fig:1} and \ref{fig:2}, the value of the
superconducting gap is smaller and therefore the values of the critical
densities for the 2SC case are closer to the ones of normal quark
matter and the unpairing transition occurs at larger values of the
proton fraction. This effect is evident in Fig.~\ref{fig:1}, for intermediate diquark pairing,
but is not so pronounced in Fig.~\ref{fig:2} where the strong diquark pairing is considered
and the critical temperature is higher than $50$ MeV.

\begin{figure}
    \begin{centering}
\epsfig{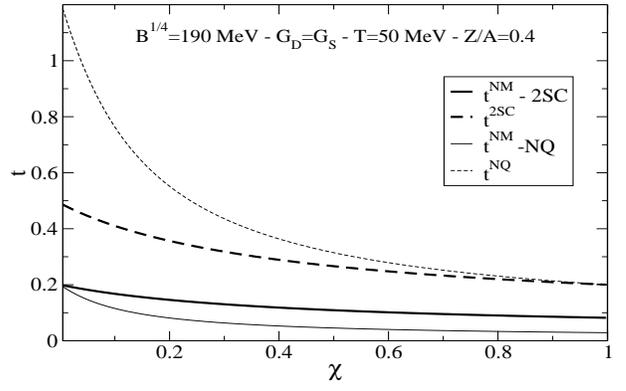}
    \caption{Local isospin asymmetries of the nuclear and quark components of the mixed phase. In the case of normal quark matter,
close to the onset of the mixed phase, the isospin asymmetry is distillated into the quark component for which $t \sim 1$. 
In the 2SC case, due to its large symmetry energy, no strong asymmetries are reached. The neutron distillation
effect is suppressed.  
 \label{asy} }
   \end{centering}
\end{figure}

Let us have a closer look now into the mixed phase itself.  Notice
that for symmetric matter, the system is actually a one component
system and therefore the Gibbs construction coincides with the Maxwell
construction. Thus, within the mixed phase the pressure is constant
and the two phases are both symmetric.  The situation is different for
asymmetric matter: as a result of the Gibbs conditions the two phases
in the mixed phase have different isospin asymmetries, the nuclear
phase being the most symmetric, and the pressure increases with the
density.  In Fig.~\ref{asy}, we show the local isospin asymmetry $t$
of the nuclear phase and the normal quark and 2SC phase within the
mixed phase as functions of the volume fraction $\chi$.  Parameters
are: $B^{1/4}=190$ MeV, $T=50$ MeV, $Z/A=0.4$, which implies $t=0.2$,
and for the 2SC phase we consider the case of strong coupling.  An
interesting qualitative difference is evident whether the quark component
is in the unpaired quark state or in the 2SC state. In the case of normal quark matter,
due to its low value of the symmetry energy, close to the onset of
the mixed phase the quark component is very asymmetric, $t \sim 1$. Notice that an asymmetry larger than $1$, which would imply 
a negative value for Z/A, is possible in pure quark matter by considering the fact that the density of protons in quark matter
is given by $n_p=(2 n_u - n_d)/3$ as can be easily verified.
On the other hand, the nuclear component, which at the onset of the mixed phase has
asymmetry $t \sim 0.2 $, becomes more and more symmetric as $\chi$
increases. This is the so called neutron distillation effect
\cite{DiToro:2006pq,DiToro:2009ig}: there is an excess of isospin
density in the quark drops with respect to the nuclear phase which,
due to its high value of the symmetry energy, lowers its energy
approaching the symmetric state.  As the volume fraction increases,
the asymmetry of the quark phase rapidly decreases and reaches the
value $t \sim 0.2 $ at $\chi=1$ as it must be.  Notice that at
$\chi=1$, the end of the mixed phase, the asymmetry of the nuclear
phase is almost zero.   
\vskip 1cm
The situation is different in the case of the
2SC phase.  Let us consider first the onset of the mixed phase: the
asymmetry in the chemical potentials of the nuclear phase affects only
the unpaired blue quarks: due to the pairing, the other quarks do not
contribute to the asymmetry. This explains why close to the onset of
the mixed phase the 2SC phase is actually much less asymmetric than
the normal quark phase. As $\chi$ increases, slowly the 2SC phase reduces
its asymmetry until the value $t=0.2$ is reached at the end of the
mixed phase; at the same time the nuclear phase reduces its asymmetry
and reaches, at $\chi=1$, a value $t\sim 0.1$, thus larger with
respect to the case of normal quark matter. This is again explained by
considering that for a fixed value of $t$ in quark matter, the
mismatch of chemical potentials of up and down quarks is larger in the
2SC phase than the normal quark phase, therefore at the end of the
mixed phase this produces a larger asymmetry of the nuclear phase when
the 2SC phase is considered.  In conclusion, the isospin distillation
effect in presence of the 2SC phase is strongly reduced with respect
to the case of normal quark matter. We will discuss in the last section
possible effects for heavy ion collisions experiments.

\begin{figure}
    \begin{centering}
\epsfig{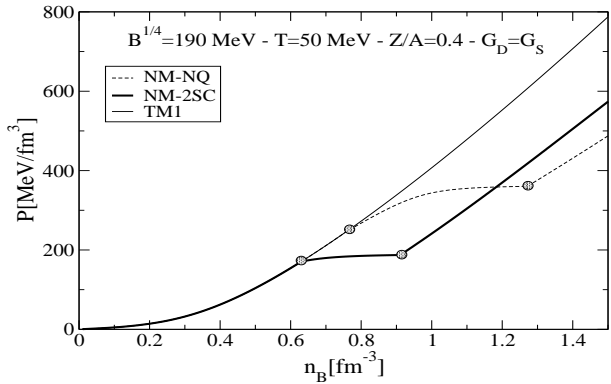}
    \caption{Equations of state, pressure as a function of the density,
           for nuclear matter TM1, normal quark matter and 2SC phase. The
dots indicate the onset and the end of the mixed phases. The softening 
of the equation of state due to the formation of 2SC matter is more pronounced then the one
of the normal quark phase.
\label{press} }
   \end{centering}
\end{figure}

In Fig.~\ref{press}, we show a comparison between the equations of
state when the normal quark phase or the 2SC phase are considered, for
the same choice of parameters. As already noticed, the onset of the
mixed phase occurs at lower densities when the 2SC phase is considered
because it is softer than the normal quark phase. One can notice that
the extension of the NM-2SC mixed phase is reduced with respect to the
case of the NM-NQ mixed phase. At the same time the variation of the
pressure in the NM-2SC mixed phase is much smaller than the one in the
NM-NQ mixed phase. Indeed the Gibbs construction in the case of the 2SC phase
provides a result which is quite similar to the simpler Maxwell
construction. Again, this is clear if we consider that the 2SC phase
has a larger symmetry energy with respect to the normal quark phase
and therefore the two components of the mixed phase both prefer to be
in a state as more symmetric as possible (similarly in neutron star
matter a phase transition from nuclear matter to the CFL phase is
treated with a Maxwell construction since the pairing in the CFL phase
already enforces its charge neutrality \cite{Alford:2001zr}). In
conclusion, the NM-2SC mixed phase is more compressible than the NM-NQ
mixed phase and, as we will discuss in the last section, this might also be
important in heavy ions collisions experiments.


\section{Phase diagrams}

\begin{figure}[t]
    \begin{centering}
\epsfig{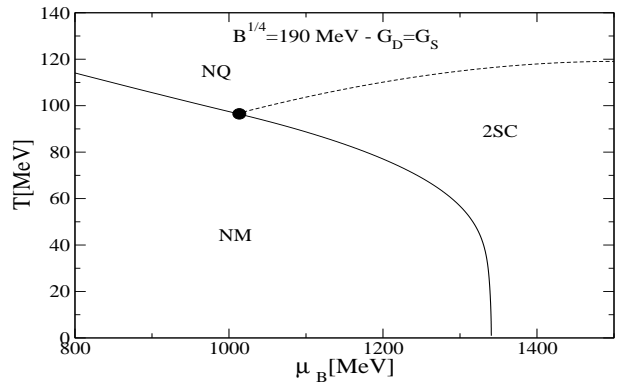}
    \caption{Phase diagram for symmetric matter in the temperature chemical potential plane.
Solid lines indicate first order phase transitions, the dashed line corresponds to the 
second order phase transition between the 2SC phase and normal quark matter. 
\label{fig:0} }
   \end{centering}
\end{figure}

In the last section we have shown how, as the proton fraction decreases,
one goes from the phase transition to the 2SC phase to the phase
transition to the normal quark phase due to the breaking of the
pairing pattern. We show in this section the effect of the
temperature: in general a second order phase transition is obtained
from the 2SC to the normal phase as the temperature increases and at fixed chemical potential, with
critical temperature $T_{crit}=0.57 \Delta_0$, where $\Delta_0$ is the
gap as obtained at zero temperature. Let us consider now 
the phase diagrams at fixed values of Z/A.  In Fig.~\ref{fig:0}, we
show the phase diagram in the temperature baryon chemical potential
plane, for strong coupling and symmetric matter. A first order
phase transition line separates nuclear matter from quark matter both
in the normal state and the 2SC state. A second order phase
transition line separates the normal quark phase from the 2SC phase.
The two lines intersect in a point at $T \sim 95$ MeV and $\mu_B \sim
1000$ MeV, thus potentially interesting for heavy ions collision
experiments as we will discuss in the following \footnote{Notice that at zero temperature the phase transition to quark matter
occurs at a chemical potential of the order of $1350$ MeV thus larger than the chemical potential of the chiral phase transition as obtained 
in the NJL-like model of Refs.~\cite{Huang:2002zd,Huang:2003xd}. This justifies our assumption to neglect here the effective mass of quarks.}.  
Interestingly, the overall
structure of the phase diagram is reminiscent of the phase diagram of
$^4$He for which at small temperatures a first order phase transition
line separates the solid phase and the two liquid phases HeI and
HeII. The two liquid phase are among them separated by a second order
phase transition line, the so called $\lambda$-line \cite{libro}.

\begin{figure}
    \begin{centering}
\epsfig{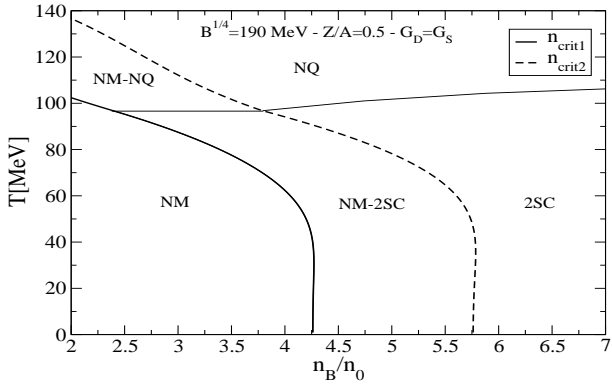}
    \caption{Phase diagram for symmetric matter in the temperature density plane.
Solid thick lines indicate the onset of the mixed phase $n_{crit1}$, dashed thick lines correspond to the end of the mixed phase $n_{crit2}$.
The solid thin line is a second order phase transition between the 2SC phase and normal quark matter. 
\label{fig:3} }
   \end{centering}
\end{figure}

\begin{figure}
    \begin{centering}
\epsfig{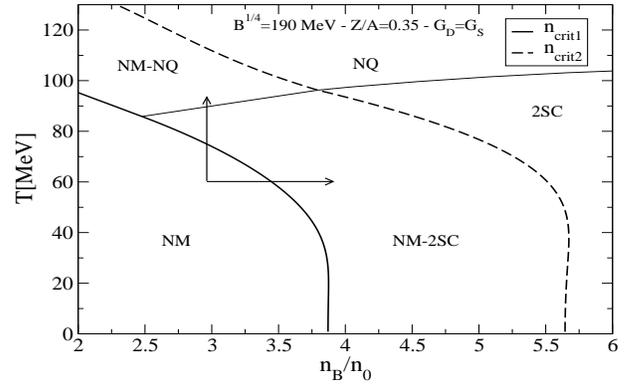}
    \caption{Phase diagram for asymmetric matter, $Z/A=0.35$, in the temperature density plane.
Solid thick lines indicate the onset of the mixed phase $n_{crit1}$, dashed thick lines correspond to the end of the mixed phase $n_{crit2}$.
The solid thin line is a second order phase transition between the 2SC phase and normal quark matter. The two arrows
indicate the region of the phase diagram which might be reached in semi-central low energy heavy ions collisions \cite{DiToro:2006pq}.
\label{fig:4} }
   \end{centering}
\end{figure}

Let us look now at the phase diagrams in the temperature density
plane. In Fig.~\ref{fig:3}, we show the case of symmetric matter and
strong diquark coupling.  Notice that the phase diagram is
divided in five regions: depending on the density and temperature,
three pure phases and two mixed phases can be formed, separated by first
order transition lines (thick dashed and solid lines) and second order
transition lines (thin solid).  To make contact with heavy ions
physics, where nuclei with $Z/A \sim 0.4$ are used, we compute the
phase diagram for asymmetric matter.  We want to discuss our phase
diagram, in comparison with the results obtained in
Ref.~\cite{DiToro:2006pq}. In that paper a transport model is used to
investigate the conditions reached in semi-central heavy ion
collisions of $^{238} U$ ($Z/A=0.387$) nuclei at $1$ A GeV.
Interestingly, it was found that rather exotic nuclear matter is
formed in a transient time of $10$ fm/c having densities around $3 n_0$, $T \sim 50-60$ MeV and $Z/A \sim 0.35-0.4$.  In
Fig.\ref{fig:4} we show the phase diagram for $Z/A =0.35 $ in the case of
strong diquark pairing.  As before, three pure phases and two mixed
phases are obtained separated by first order and second order
transition lines. Notice that within the NM-2SC mixed phase, the line
of second order phase transition to the NM-NQ mixed phase is not constant anymore
because for asymmetric matter the chemical potential varies
in the mixed phase and therefore also the superconducting gap. As we explained before, 
in the case of asymmetric matter the onset of the phase transition
occurs at lower densities with respect to the case of symmetric matter
and the  window of mixed phase is larger.
Finally, the two
arrows in Fig.~\ref{fig:4} indicate the region of the phase diagram which can be reached
in experiments as proposed in Ref.~\cite{DiToro:2006pq}.
Clearly, under the assumption of strong diquark pairing, which implies 
a critical temperature for color superconductivity of $\sim 80-100$ MeV, it
would be possible to reach the NM-2SC mixed phase in heavy ions
collisions.

\section{Discussion and Conclusions}
We have computed the mixed phase of nuclear matter and 2SC phase
under different conditions of temperature and isospin asymmetry (or
proton fraction).  First, we have provided a clear argument by which
the symmetry energy of the 2SC phase is, to a good approximation,
three times larger than the one of normal quark matter: due to the Cooper
pairing of red and green quarks, only the blue unpaired up and down
quarks in the 2SC phase can contribute to the isospin asymmetry. The
argument is confirmed by a numerical calculation. This fact allows to
understand the main differences between the nuclear matter 2SC mixed
phase and the one with the normal quark phase: the onset of
the mixed phase shows a mild dependence on Z/A because both the nuclear and the 2SC 
equations of state become substantially stiffer when the proton fraction is reduced.
The isospin distillation effect proposed in \cite{DiToro:2006pq} in the case of normal 
quark matter is reduced in the nuclear matter 2SC mixed phase because of the 
larger value of its symmetry energy. Finally, the softening of the equation of state
due to the appearance of the nuclear matter quark matter mixed phase is more pronounced 
in the case of the 2SC phase with respect to the case of normal quark matter.
A further interesting point is that the phase transition can occur 
at lower densities than previously thought when 
the effects from color superconductivity are taken into account.
The crucial question is whether these differences can be
probed in heavy ions collisions experiments. By referring to the
results of the transport calculation of \cite{DiToro:2006pq} it might be
possible indeed to enter in the nuclear matter 2SC mixed phase in low
energy semi-central collisions. Concerning the possible observables
which allow to distinguish whether normal quark phase or the 2SC phase
are formed, here we want to limit ourself to discuss qualitatively a
few ideas.

The fact that in a mixed phase between two different components,
nuclear matter and quark matter, there can be a separation or
distillation of globally conserved quantities from one phase to the
other is rather old.  It was indeed proposed in
Ref.~\cite{Greiner:1987tg}, that in high energy heavy ions collisions
strange and antistrange quarks are abundantly produced (with net
strangeness being zero) and that strangeness would be much more
abundant in the quark component.  This could be a mechanism to
produce strangelets (stable or metastable) in the
laboratories. Similarly, the isospin distillation effect proposed in
\cite{DiToro:2006pq} is the migration of the isospin density into the
phase with the lowest value of the symmetry energy. It has been
proposed that this phenomenon could invert the trend in the production
of neutron rich fragments and it could affect the $\pi^-/\pi^+$
multiplicities ratio.  In the first studies of color superconductivity
it was argued that due to the formation of Cooper pairs in the densest
region of the system, the quark phase would expel the excess of down
quarks and up antiquarks, which then, in the hadronic phase, would
eventually form $\pi^-$. Moreover, when the diquark condensate breaks up
late in the collision a number of protons larger then the initial
one in the colliding nuclei would be emitted.  The signature would
then be an increase of the $\pi^-/\pi^+$ ratio and at the same time an
increase of the number of protons within selected events with
an anomalously large density and small temperature
\cite{Alford:1997ps,Kiriyama:2001ud}. This effect could be regarded as
the opposite of the neutron distillation effect, the quark phase is
completely symmetric and expels its excess of isospin into the nuclear
phase.  This would be correct if all the quarks pair, but in the 2SC
phase the unpaired up and down blue quarks can, as in the normal quark
phase, have different densities.  Actually the symmetry energy of the
2SC phase, while larger then the one of the normal phase, is still
smaller than the nuclear matter symmetry energy (at least within the model of nuclear matter we consider here).  
Therefore as we have shown, also when considering the 2SC phase a neutron distillation into
the quark component of the mixed phase occurs but it is simply
suppressed with respect to the normal quark phase.  One could then
expect that, going from low temperatures at which the 2SC phase is
formed to higher temperatures (by increasing the energy of the ions)
where the normal phase is formed, the neutron distillation effects is
gradually enhanced and therfore also its specific signatures
\cite{DiToro:2006pq}.  At the same time, the mixed phase becomes
stiffer passing from the 2SC phase to the normal quark phase what can
be for instance "detected" by using the $K^+$ yields which were shown
to represent a good probe for the stiffness and the isospin dependence of the equation of state
\cite{Aichelin:1986ss,Sturm:2000dm,Fuchs:2000kp,Hartnack:2005tr,Ferini:2006je,Forster:2007qk,Sagert:2007nt}.  
Other possible signatures of the formation of quark matter are associated 
with the enhancement of the $\bar{\Lambda}$ to $\bar{p}$ ratio as shown in Refs.~\cite{Stephans:1997qi,Armstrong:1998ih,Back:2001ai} although alternative explanations based on multihadron reactions have been proposed \cite{Rapp:2000gy,Greiner:2000tu}.
A detailed quantitative study of
these quantities within a transport model would be of course very important.
In addition to the particles yields one can calculate also the
susceptibilites as done in Refs.~\cite{Sasaki:2006ws,Sasaki:2006ww}
which are important for the charge (baryonic, electric, isospin)
fluctuations.  In particular one could expect that the off-diagonal
susceptibility $\chi_{ud}$ could be different in the 2SC phase due to
the Cooper pair correlations. Finally, the results of this paper
might be relevant also for neutron stars physics: in protoneutron
stars, where the initial proton fraction is $\sim 0.3$, during
deleptonization the proton fraction decreases and at some point it
might be possible that a phase transition from the 2SC phase to the
normal quark phase can take place due to the increasing stress on the
quark's Fermi surfaces \cite{Ruester:2005ib,Sandin:2007zr,Pagliara:2007ph}. Moreover,
the local proton fraction of the nuclear phase within the mixed phase
is also important for the late cooling of neutron stars since it
regulates the threshold of the direct Urca processes \cite{Page:2005fq}.

The work of G.~P. is supported by the Deutsche Forschungsgemeinschaft
(DFG) under Grant No. PA 1780/2-1. J.~S.~B. is supported by the DFG
through the Heidelberg Graduate School of Fundamental Physics.  We
thank A.~Drago and M.~Hempel for many fruitful discussions. This work
was also supported by CompStar, a Research Networking Programme of the
European Science Foundation.

\bibliography{references}
\bibliographystyle{apsrev}

\end{document}